\documentclass{article}%
\usepackage{amsmath}
\usepackage{amsfonts}
\usepackage{amssymb}
\usepackage{graphicx}%
\providecommand{\U}[1]{\protect\rule{.1in}{.1in}}

\begin{document}

\title{What is ontic and what is epistemic in the Quantum Mechanics of Spin?}
\author{Ariel Caticha\\{\small Department of Physics, University at Albany-SUNY, }\\{\small Albany, NY 12222, USA.}}
\date{}
\maketitle

\begin{abstract}
Entropic Dynamics (ED) provides a framework that allows the reconstruction of
the quantum formalism by insisting on ontological and epistemic clarity and
adopting entropic methods and information geometry. Our present goal is to
extend the ED framework to account for spin. The result is a \emph{realist
}$\psi$\emph{-epistemic model} in which the ontology consists of a particle
described by a definite position plus a discrete variable that describes
Pauli's peculiar two-valuedness.\ The resulting dynamics of probabilities is,
as might be expected, described by the Pauli equation. What may be unexpected
is that the generators of transformations -- Hamiltonians and angular momenta
including spin, are all granted clear epistemic status. To the old question
`what is spinning?' ED provides a crisp answer: \emph{nothing is spinning}.

\end{abstract}

\section{Introduction}

\label{intro}The framework of Entropic Dynamics (ED) allows the formulation of
dynamical theories as an application of the method of maximum entropy
\cite{Jaynes 1983}\cite{Jaynes 2003}\cite{Caticha 2024}. It allows, among
other things, a derivation of the Schr\"{o}dinger equation including its
linear and complex structure, and it also clarifies the most controversial
aspect of quantum mechanics (QM) --- its interpretation \cite{Caticha
2024}\cite{Caticha 2019}\cite{Caticha 2021}. The controversy centers around
the question of what is real or \emph{ontic} at the microscopic quantum level.
Furthermore, it is not clear either how the \emph{epistemic} aspects of the
theory are handled. Are probabilities already present at the microscopic level
or do they only arise at the macroscopic classical level when measurements are
performed? Could we perhaps need some altogether different type of quantum or
exotic probability?

In the ED approach the main concern is to achieve ontological and
epistemological clarity and, therefore, before any further discussion it may
be desirable to be explicit about the terminology: The paradigmatic ontic
concept is matter; a quantity is said to be `ontic' when it refers to
something real, substantial. The paradigmatic epistemic quantities are
probabilities and wave functions; a concept is `epistemic' when it refers to
the state of belief, opinion, or knowledge of an agent (who, for our current
purposes, we shall assume to be an ideally rational agent). Models such as ED
that invoke ontic variables (\emph{e.g.}, position) while the wave function
remains fully epistemic are often described as \textquotedblleft realist
$\psi$-epistemic models.\textquotedblright\ There exist powerful no-go
theorems that rule out large families of such models --- the so-called
\textquotedblleft ontological\textquotedblright\ models --- because they
disagree with QM. For an extended list of references and a discussion of how
ED evades those no-go theorems see \cite{Caticha 2022}. Briefly, ED is realist
and $\psi$-epistemic, but it is not an ontological model.

On a related issue, it is important to emphasize that the distinction
ontic/epistemic is not the same as the distinction objective/subjective.
Probabilities, for example, are always fully epistemic (because they codify
credences or degrees of belief) but they can lie anywhere in the spectrum from
objective to subjective. To be explicit: probabilities in QM are fully
objective, but in other contexts probabilities can be subjective because two
agents could hold different beliefs as a result of different priors or
different data. This paper, however, is not about philosophy; it is about
extending the ED framework to the discussion of the quantum mechanics of spin.

In 1923 Pauli introduced a \textquotedblleft peculiar, classically not
describable two-valuedness\textquotedblright\ \cite{Pauli 1946}\cite{Jammer
1966} associated to spin and to the exclusion principle. Spin is an angular
momentum and that, in itself, is not particularly strange, but ever since
Uhlenbeck and Goudsmit's 1925 idea of spin \cite{Uhlenbeck Goudsmit 1925}, the
nature of what, if anything, is actually spinning, and the nature of the
peculiar two-valuedness has been and continues to be a subject of interest and
of numerous studies.

The standard Copenhagen interpretation is silent on these questions; it
forgoes the possibility of any visualizable spin models. In other
interpretations, however, spin can be an ontic variable that is variously
attributed either to a spinning particle, or to the helical motion of a point
particle as guided by a wave function, and the latter might be a Bohmian wave
function, a Pauli spinor, a Dirac spinor, a real spinor function in the
language of geometric algebra, or stochastic mechanics, or more closely
related to quantum information. A non-exhaustive list of references includes
\cite{Bohm et al 1955}-\cite{Powers et al 2025}; in particular, \cite{Ohanian
1986} and \cite{Hestenes 1990} deal with the interpretation of relativistic
spin and the Dirac equation, and \cite{Budiyono Rohrlich 2017}\thinspace
\cite{Beyer Paul 2023}\thinspace\cite{Powers et al 2025} contain a wealth of
recent references, both to foundational and more computational applications.
None of these models, however, provide insights as to why the Pauli equation
takes the particular form it does (\emph{e.g.}, how does one derive the
linearity, the adoption of complex numbers, and so on).

In a previous work \cite{Caticha Carrara 2020}\cite{Carrara 2021} we presented
a nonrelativistic ED model for a single spin-1/2 point particle. The position
of the particle was assumed to be the \emph{only ontic variable} and spin was
recovered as a \emph{property of its epistemic wave function}. The model was
successful in the sense that it provided a reconstruction of the
single-particle Pauli equation but two features have hindered its satisfactory
extension to several particles. The first is that the four real degrees of
freedom of the single particle spinor allow an elegant geometric
interpretation in terms of a probability density plus the three Euler angles
that define the rotation from a fixed lab frame to a spatially varying
\textquotedblleft spin\textquotedblright\ frame attached to the particle
\cite{Caticha Carrara 2020}. Unfortunately, this very appealing feature does
not generalize to several particles because it is not in general possible to
attach a separately rotating spin frame to each individual particle. One
specific goal of the present paper is to provide an ED reconstruction of the
Pauli equation that could in principle be extended to several particles
because it does not rely on individual spin frames.

The second feature is relevant to the eventual extension to identical
particles. The purpose of deploying an information-based framework such as ED
to reconstruct QM is to provide natural explanations for the typical quantum
effects --- interference, entanglement, tunneling, etc. The Pauli exclusion
principle, however, has so far proved resistant in the sense that it could
only be implemented by force, that is, by an ad hoc, unexplained
antisymmetrization of the wave function. Here we take a first step towards a
more natural explanation.

It turns out that, among all quantum effects, the exclusion principle is
unique in that it is remarkably robust. While effects such as interference,
entanglement, tunneling, etc., are all destroyed by noise, the exclusion
principle can survive under the most extreme conditions such as, for example,
in the interior of stars. A natural explanation would follow from the
observation that quantum effects that are sensitive to noise and decoherence
can all be traced to\ the wave function, that is, to the epistemic sector of
ED. The robustness of the exclusion principle strongly suggests that its
explanation lies in the ontic sector.

We shall reconstruct the non-relativistic one-particle Pauli equation by
enlarging the ontology to include both the position of the point particle and
Pauli's discrete two-valued variable. This answers the question `what is
real?' and defines what variables we are uncertain about (Section
\ref{ontology}). The probabilities of these variables form a statistical
manifold --- the epistemic configuration space --- and its associated
cotangent bundle constitutes the epistemic phase space. Next, in Section
\ref{HK flows}, we briefly discuss the kinematics of Hamilton-Killing flows,
which singles out those special curves that are adapted to the natural
geometrical structures of the epistemic phase space. The discussion of which
among those special curves qualify to describe evolution in time --- this is
the actual entropic dynamics --- starts in Section \ref{short steps} where we
study the ED of infinitesimally short steps, followed by the construction of
an entropic notion of time in Section \ref{entropic time}, and the derivation
of a continuity equation for the evolution for probability in Section
\ref{Probability evolution}. In Section \ref{Hamiltonian} we derive the
corresponding Hamiltonian and the Pauli equation. The reconstruction of
orbital and spin angular momenta as generators of those Hamilton-Killing flows
that also generate rotations is given in Section \ref{angular momentum}. Some
final thoughts and conclusions are collected in Section \ref{conclusions}. To
make this paper somewhat self-contained some material presented in
\cite{Caticha 2019} is reproduced here. However, the present paper reflects
substantial differences: it derives subquantum trajectories that are
non-differentiable and Brownian, while in \cite{Caticha 2019} they are smooth
like Bohmian trajectories; furthermore, the addition of a discrete two-valued
variable requires a revised treatment of time.

\section{The ontic and the epistemic sectors}

\label{ontology}The first step is to specify the subject matter --- the
ontology. We consider a point particle living in a flat Euclidean space. The
particle is assumed to have a \emph{definite }position described as
$x=\{x^{a},~a=1,2,3\}$ in Cartesian coordinates. In addition we assume the
particle occupies a definite state denoted $k=\{-1,+1\}$, which corresponds to
Pauli's peculiar, \textquotedblleft classically not describable
two-valuedness\textquotedblright. Both the assumption of a definite $x$ and of
a definite $k$ already represent a major departure from the standard
Copenhagen interpretation.

Next, we discuss the epistemic sector. ED is a dynamics of probabilities. The
goal is to study the evolution of the joint probability distribution
$\rho(k,x)$ and its canonically conjugate momentum $\xi(k,x)$. We adopt the
following notation: we shall often abbreviate $k=\pm1$ by $k=\pm$ and write
\begin{equation}
\rho(k,x)=\rho_{kx}=\rho_{\pm x}\quad\text{and}\quad\xi(k,x)=\xi_{kx}=\xi_{\pm
x}~. \label{rho csi}%
\end{equation}
As discussed in \cite{Caticha 2024}\cite{Caticha 2019}\cite{Caticha 2021} it
is convenient to transform from the generalized coordinates $(\rho,\xi)$ to
complex coordinates, known as the wave function,
\begin{equation}
\Psi(x)=\binom{\psi_{+x}}{\psi_{-x}}\quad\text{where}\quad\psi_{\pm x}%
=\psi_{kx}=\rho_{kx}^{1/2}\exp\frac{i}{\hbar}\xi_{kx}~. \label{spinor a}%
\end{equation}
Clearly, the wave function $\Psi(x)$ also belongs in the epistemic sector. The
new canonically conjugate pairs are $(\psi_{kx},i\hbar\psi_{kx}^{\ast})$ and
the transformation $(\rho_{kx},\xi_{kx})\rightarrow(\psi_{kx},i\hbar\psi
_{kx}^{\ast})$ is a canonical transformation. The generators of translations
and rotations, the momentum $\tilde{P}[\rho,\xi]$ and the angular momentum
$\tilde{J}[\rho,\xi]$, are quantities that obviously also belong in the
epistemic sector.

The weight of tradition leads us to refer to $k$ as `spin' but this might not
be fully appropriate because $k$ is ontic while spin, being an angular
momentum, is an epistemic quantity. We might also refer to $k$ as a `qubit'
but some caution, however, is called for because of the need to distinguish an
ontic qubit (a discrete two-valued ontic variable like $k$) from an epistemic
qubit (a two-dimensional Hilbert space). It seems clear that talking about
spin or about qubits without being aware of what is ontic and what is
epistemic in QM will lead to considerable confusion. Interestingly, the same
kind of confusion can arise with the classical term `bit' which applies both
to the ontic bit (which refers to binary subsystem in the physical memory of a
digital computer) and to the epistemic bit (which refers to a unit of amount
of information as measured by the Shannon entropy). This kind of confusion is
often reflected in expressions such as \textquotedblleft information is
physical\textquotedblright. Nevertheless, tradition weighs heavy, and we shall
often use the terms `spin' or `qubit' and trust that whether one refers to the
ontic or the epistemic version can be understood from the context.

\section{Kinematics:\ Hamilton-Killing flows}

\label{HK flows}The discussion of Hamilton-Killing or HK flows can be carried
out by a straightforward extension of the treatment for discrete variables
(\emph{e.g.}, the quantum die) in \cite{Caticha 2021} and for continuous
variables (particle positions) in \cite{Caticha 2019}. Here we shall omit most
technical details; for a pedagogical discussion see \cite{Caticha 2024}.

Once local coordinates $\{\rho_{kx},\xi_{kx}\}$ on the e-phase space, have
been established there is a natural choice of symplectic form,
\begin{equation}
\Omega\overset{\text{def}}{=}\int dx%
{\textstyle\sum\limits_{k}}
\left(  \,\tilde{\nabla}\rho_{kx}\otimes\tilde{\nabla}\xi_{kx}-\tilde{\nabla
}\xi_{kx}\otimes\tilde{\nabla}\rho_{kx}\right)  ~, \label{sympl form b}%
\end{equation}
where $\,\tilde{\nabla}$ is the gradient in e-phase space. Alternatively, we
can do a canonical transformation to complex coordinates $\{\psi_{kx}%
,i\hbar\psi_{kx}^{\ast}\}$, eq.(\ref{spinor a}), and let the coordinates of a
point $\Psi$ in e-phase space be
\begin{equation}
\Psi^{\mu x}=\left(  \Psi^{1x},\Psi^{2x},\Psi^{3x},\Psi^{4x}\right)  =\left(
\psi_{+x},i\hbar\psi_{+x}^{\ast},\psi_{-x},i\hbar\psi_{-x}^{\ast}\right)  ~,
\label{Psi mu}%
\end{equation}
then
\begin{equation}
\Omega\overset{\text{def}}{=}\int dx%
{\textstyle\sum\limits_{k}}
\left(  \,\tilde{\nabla}\psi_{kx}\otimes\tilde{\nabla}i\hbar\psi_{kx}^{\ast
}-\tilde{\nabla}i\hbar\psi_{kx}^{\ast}\otimes\tilde{\nabla}\psi_{kx}\right)
~, \label{sympl form a}%
\end{equation}
and the tensor components of $\Omega$ are
\begin{equation}
\lbrack\Omega_{\mu x,\mu^{\prime}x^{\prime}}]=%
\begin{bmatrix}
0 & 1 & 0 & 0\\
-1 & 0 & 0 & 0\\
0 & 0 & 0 & 1\\
0 & 0 & -1 & 0
\end{bmatrix}
\,\delta(x,x^{\prime})~. \label{sympl form c}%
\end{equation}
(Notation: $[\Omega_{\mu x,\mu^{\prime}x^{\prime}}]$ is a $4\times4$ matrix
the elements of which, $\Omega_{\mu\mu^{\prime}}$, are functions of $x$ and
$x^{\prime}$.)

Consider a curve $\Psi^{\mu x}(\tau)$ on the e-phase space parametrized by
$\tau$ and let
\begin{equation}
\bar{H}=\frac{d}{d\tau}=H^{\mu x}\frac{\delta}{\delta\Psi^{\mu x}}%
\quad\text{with}\quad H^{\mu x}[\Psi]=\frac{d\Psi^{\mu x}}{d\tau}~
\label{Ham Vector}%
\end{equation}
be the tangent vector at $\Psi$. We are interested in those special curves
that are naturally adapted to the symplectic geometry in the sense that they
preserve $\Omega$, that is,
\begin{equation}
\pounds _{\bar{H}}\Omega=0~,
\end{equation}
where $\pounds _{\bar{H}}$ is the Lie derivative along $\bar{H}[\Psi]$ (this
is a directional derivative on a curved space). By Poincare's lemma, requiring
$\pounds _{\bar{H}}\Omega=0$ (a vanishing \textquotedblleft
curl\textquotedblright) implies that the covector $\Omega_{\mu x,\mu^{\prime
}x^{\prime}}H^{\mu^{\prime}x^{\prime}}$ is the gradient of a scalar function
\cite{Schutz 1980}, denoted $\tilde{H}[\Psi]$,
\begin{equation}
\Omega_{\mu x,\mu^{\prime}x^{\prime}}H^{\mu^{\prime}x^{\prime}}=\frac{\delta
}{\delta\Psi^{\mu x}}\tilde{H}[\Psi]~. \label{Ham grad}%
\end{equation}
($\tilde{H}[\Psi]$ is the scalar function associated to the vector $\bar
{H}[\Psi]$.) Substituting (\ref{sympl form c}) and (\ref{Ham Vector}) this is
rewritten as
\begin{equation}
\frac{d\psi_{kx}}{d\tau}=\frac{\delta\tilde{H}}{\delta(i\hbar\psi_{kx}^{\ast
})}\quad\text{and}\quad\frac{d(i\hbar\psi_{kx}^{\ast})}{d\tau}=-\frac
{\delta\tilde{H}}{\delta\psi_{kx}}~, \label{Hamiltonian flow a}%
\end{equation}
which are recognized as Hamilton's equations for a Hamiltonian function
$\tilde{H}$. This is the reason for Hamiltonians in physics: the congruence of
curves that preserve the natural symplectic geometry, $\pounds _{\bar{H}%
}\Omega=0$ are called \emph{Hamilton flows}. They are generated by
\emph{Hamiltonian vector fields} $\bar{H}$ or, equivalently, by their
associated \emph{Hamiltonian functions} $\tilde{H}$. Our challenge, to be
addressed next, is to find Hamiltonians $\tilde{H}$ that yield interesting flows.

It turns out that in addition to the symplectic geometry $\Omega$, the e-phase
also has a natural metric geometry inherited from the information geometry of
the e-configuration space, which is a statistical manifold. A straightforward
extension from \cite{Caticha 2024}\cite{Caticha 2019}\cite{Caticha 2021}
yields a particularly simple line element,
\begin{equation}
\delta\ell^{2}=\int dxdx^{\prime}%
{\textstyle\sum\limits_{\mu\mu^{\prime}}}
G_{\mu x,\mu^{\prime}x^{\prime}}\delta\Psi^{\mu x}\delta\Psi^{\mu^{\prime
}x^{\prime}}=2\hbar\int dx%
{\textstyle\sum\limits_{k}}
\,\delta\psi_{kx}\delta\psi_{kx}^{\ast}~. \label{spinor metric c}%
\end{equation}
The components of the metric tensor $G$ and its inverse are
\begin{equation}
\lbrack G_{\mu x,\mu^{\prime}x^{\prime}}]=-i%
\begin{bmatrix}
0 & 1 & 0 & 0\\
1 & 0 & 0 & 0\\
0 & 0 & 0 & 1\\
0 & 0 & 1 & 0
\end{bmatrix}
\,\delta(x,x^{\prime})~,\quad\lbrack G^{\mu x,\mu^{\prime}x^{\prime}}]=i%
\begin{bmatrix}
0 & 1 & 0 & 0\\
1 & 0 & 0 & 0\\
0 & 0 & 0 & 1\\
0 & 0 & 1 & 0
\end{bmatrix}
\,\delta(x,x^{\prime}). \label{spinor metric e}%
\end{equation}
A remarkable further development is that the contraction of the symplectic
form $\Omega$, eq.(\ref{sympl form c}), with the inverse metric $G^{-1}$
allows us to construct a tensor $J$ with components
\begin{equation}
J^{\mu x}{}_{\mu^{\prime}x^{\prime}}=-\int dx^{\prime\prime}%
{\textstyle\sum\limits_{\mu^{\prime\prime}}}
G^{\mu x,\mu^{\prime\prime}x^{\prime\prime}}\Omega_{\mu^{\prime\prime
}x^{\prime\prime},\mu x^{\prime}}\quad\text{and}\quad\lbrack J^{\mu x}{}%
_{\mu^{\prime}x^{\prime}}]=%
\begin{bmatrix}
i & 0 & 0 & 0\\
0 & -i & 0 & 0\\
0 & 0 & i & 0\\
0 & 0 & 0 & -i
\end{bmatrix}
\,\delta(x,x^{\prime})~. \label{J tensor}%
\end{equation}
What makes the tensor $J$ special is that its square is
\begin{equation}
\int dx^{\prime\prime}%
{\textstyle\sum\limits_{\mu^{\prime\prime}}}
J^{\mu x}{}_{\mu^{\prime\prime}x^{\prime\prime}}J^{\mu^{\prime\prime}%
x^{\prime\prime}}{}_{\mu^{\prime}x^{\prime}}=-\delta^{\mu}{}_{\mu^{\prime}%
}\delta(x,x^{\prime})~.
\end{equation}
In words, the action of $J^{2}$ is equivalent to multiplying by $-1$, which
means that $J$ provides a complex structure. This is the reason for complex
numbers in QM and explains why it was convenient to introduce wave functions
(i.e., complex coordinates) in the first place.

Next, we take advantage of the metric geometry and seek those special curves
that preserve both the symplectic and the metric geometries,
\begin{equation}
\pounds _{\bar{H}}\Omega=0\quad\text{and}\quad\pounds _{\bar{H}}G=0~.
\end{equation}
We want $\bar{H}$ to be both a Hamilton and a Killing vector field;\ then the
associated Hamiltonian function $\tilde{H}$ will generate Hamilton-Killing
flows. Imposing further that the HK\ flows preserve the normalization of
probabilities restricts the Hamiltonian function to those that are bilinear in
$\psi_{kx}$ and $\psi_{kx}^{\ast}$,%

\begin{equation}
\tilde{H}=\int dxdx^{\prime}%
{\textstyle\sum\limits_{kk^{\prime}}}
\psi_{kx}^{\ast}\hat{H}_{kx,k^{\prime}x^{\prime}}\psi_{k^{\prime}x^{\prime}}~.
\label{e-hamiltonian a}%
\end{equation}
$\tilde{H}$ is the Hamiltonian that generates evolution on the e-phase space
-- the epistemic phase space -- and is, accordingly, called the e-Hamiltonian.
From eq.(\ref{Hamiltonian flow a}) the corresponding equation of evolution is
\begin{equation}
i\hbar\frac{\partial\psi_{kx}}{\partial\tau}=\frac{\delta\tilde{H}}{\delta
\psi_{kx}^{\ast}}=\int dx^{\prime}%
{\textstyle\sum\limits_{k^{\prime}}}
\hat{H}_{kx,k^{\prime}x^{\prime}}\psi_{k^{\prime}x^{\prime}}~,
\label{Sch eq a}%
\end{equation}
which is recognized as a Schr\"{o}dinger equation.

At this point in the development $\tau$ is just a parameter along a curve;
there is no implication that the curve represents time evolution and $\tau$ is
time, or that the curve is generated by rotations about an axis and $\tau$ is
a rotation angle. To identify those special curves demands that additional
information be incorporated into the analysis in order to constrain the
e-Hamiltonian function $\tilde{H}$ beyond the generic bilinear form,
eq.(\ref{e-hamiltonian a}).

\section{The Entropic Dynamics of short steps}

\label{short steps}Beyond the reconstruction of the framework of QM including
its interpretation, another central goal of ED is the formulation of an
information-based notion of time. This involves the introduction of the
concept of an instant, the notion that the instants are suitably ordered, and
a convenient definition of duration. Remarkably, by its very construction,
associated to an entropic dynamics there is a natural arrow of entropic time.

The physically relevant information that will allow us to recover a
satisfactory concept of time is that the particle follows a continuous
trajectory in space. The continuity allows the dynamics to be analyzed as a
sequence of a large number of infinitesimally short steps $kx\rightarrow
k^{\prime}x^{\prime}$. The difficulty in the presence of spin is that the term
`short' refers to a short \emph{spatial} step,
\begin{equation}
\Delta x^{a}=x^{\prime a}-x^{a}\rightarrow0~,
\end{equation}
but the $\Delta k$ steps could be discontinuous, either $\Delta k=0$
or$\ \pm1$.\ Thus, in order to incorporate the physical information that
trajectories are continuous the $k$ variables must, at least provisionally, be
removed. This is achieved by averaging over the initial $k$ and the final
$k^{\prime}$.

The evolution of the joint probability $\rho_{kx}$ is given by%
\begin{equation}
\rho_{k^{\prime}x^{\prime}}^{\prime}=\int dx%
{\textstyle\sum\nolimits_{k}}
P(k^{\prime}x^{\prime}|kx)\rho_{kx}~.
\end{equation}
Averaging over the final $k^{\prime}$ and using the product rule,
\begin{equation}
\rho_{kx}=\rho_{x}\rho_{k|x}\quad\text{where}\quad\rho_{x}=%
{\textstyle\sum\limits_{k}}
\rho_{kx}~,
\end{equation}
gives
\begin{equation}
\rho_{x^{\prime}}^{\prime}=\int dx%
{\textstyle\sum\limits_{kk^{\prime}}}
P(k^{\prime}x^{\prime}|kx)\rho_{kx}=\int dx\left[
{\textstyle\sum\limits_{kk^{\prime}}}
P(k^{\prime}x^{\prime}|kx)\rho_{k|x}\right]  \rho_{x}~. \label{spatial evol a}%
\end{equation}
Therefore, the \emph{spatial probability }$\rho_{x}$ evolves according to
\begin{equation}
\rho_{x^{\prime}}^{\prime}=\int dx\,P(x^{\prime}|x)\rho_{x}\quad
\text{where}\quad P(x^{\prime}|x)=%
{\textstyle\sum\nolimits_{kk^{\prime}}}
P(k^{\prime}x^{\prime}|kx)\rho_{k|x}~. \label{spatial evol b}%
\end{equation}

Our immediate goal is to derive the spatial transition probability
$P(x^{\prime}|x)$. Our argument closely follows the ED of scalar particles.
The transition probability $P(x^{\prime}|x)$ is found by maximizing the
entropy,
\begin{equation}
S[P,Q]=-\int dx^{\prime}\,P(x^{\prime}|x)\log\frac{P(x^{\prime}|x)}%
{Q(x^{\prime}|x)}~,~ \label{entropy a}%
\end{equation}
relative to a prior $Q(x^{\prime}|x)$ and subject to constraints that
implement the physically relevant information that we associate with quantum
evolution \cite{Jaynes 1983}\cite{Jaynes 2003}\cite{Caticha 2024}.

We require the prior $Q(x^{\prime}|x)$ to codify the physical information that
all short steps have in common, namely, they are infinitesimally short, but
$Q$ should otherwise remain maximally non-informative; it should not induce a
preferred directionality to the motion. Such a prior can itself be derived by
maximizing an entropy,
\begin{equation}
S[Q,\mu]=-\int dx^{\prime}\,Q(x^{\prime}|x)\log\frac{Q(x^{\prime}|x)}%
{\mu(x^{\prime}|x)}~,
\end{equation}
relative to some sufficiently smooth distribution $\mu(x^{\prime}|x)$ and
subject to normalization and the rotationally invariant constraint,
\begin{equation}
\langle\delta_{ab}\Delta x^{a}\Delta x^{b}\rangle=\kappa~,
\end{equation}
with the small quantity $\kappa$ to be specified below. The result is a
Gaussian distribution,
\begin{equation}
Q(x^{\prime}|x)\propto\exp-\frac{1}{2}\alpha\delta_{ab}\Delta x^{a}\Delta
x^{b}~, \label{spatial prior}%
\end{equation}
where $\alpha$ is a Lagrange multiplier. To enforce short steps we shall
eventually take the limit $\alpha\rightarrow\infty$, which amounts to taking
$\kappa\rightarrow0$. (In the $\alpha\rightarrow\infty$ limit the prior
$Q(x^{\prime}|x)$ becomes independent of the choice of the distribution
$\mu(x^{\prime}|x)$ provided it is sufficiently smooth.)

The physical information about directionality and correlations is introduced
via a \textquotedblleft phase\textquotedblright\ constraint that follows the
standard ED strategy. For \emph{scalar} particles one introduces a
\textquotedblleft drift potential\textquotedblright\ $\varphi_{x}$ that is
canonically conjugate to the probability distribution $\rho_{x}$; it obeys the
canonical Poisson brackets,
\begin{equation}
\{\rho_{x},\varphi_{x^{\prime}}\}=\delta_{xx^{\prime}}\,,\{\rho_{x}%
,\rho_{x^{\prime}}\}=0=\{\varphi_{x},\varphi_{x^{\prime}}\}~.
\end{equation}
(Eventually a canonical transformation is performed that replaces $\varphi
_{x}$ by a more convenient momentum $\xi_{x}$.) Then, the relevant dynamical
information is imposed via a constraint on the component of the expected
displacement $\langle\Delta x^{a}\rangle$ along the gradient of $\varphi$,
\begin{equation}
\langle\Delta x^{a}\rangle\partial_{a}\varphi=\kappa^{\prime}.
\end{equation}
Here, to account for spin, we introduce two drift potentials, $\varphi
_{k}(x)=\varphi_{kx}$, that are canonically conjugate to the distribution
$\rho_{kx}$,
\begin{equation}
\{\rho_{kx},\varphi_{k^{\prime}x^{\prime}}\}=\delta_{kk^{\prime}}%
\delta_{xx^{\prime}}\,,\{\rho_{kx},\rho_{k^{\prime}x^{\prime}}\}=0=\{\varphi
_{kx},\varphi_{k^{\prime}x^{\prime}}\}~. \label{PB a}%
\end{equation}
Since continuity, \emph{i.e.}, short steps, reflects the purely spatial aspect
of the transition probability $P(x^{\prime}|x)$, the phase constraint\textbf{
}takes the form
\begin{equation}
\langle\Delta x^{a}\rangle_{x}\overline{\partial_{a}\varphi}_{x}%
=\kappa^{\prime},\quad\text{where}\quad\langle\Delta x^{a}\rangle_{x}=\int
dx^{\prime}\,P(x^{\prime}|x)\,\Delta x^{a}~, \label{constraint drift}%
\end{equation}
and where the effective gradient $\overline{\partial_{a}\varphi}$ is obtained
by averaging the gradients $\partial_{a}\varphi_{kx}$ over $k$,
\begin{equation}
\overline{\partial_{a}\varphi}_{x}=%
{\textstyle\sum\limits_{k}}
\rho_{k|x}\partial_{a}\varphi_{kx}~, \label{grad phi bar}%
\end{equation}
and $\rho_{k|x}=\rho_{kx}/\rho_{x}$. The effect of interactions with an
external electromagnetic field is handled in the same way as for scalar
particles: the gauge constraint is
\begin{equation}
\langle\Delta x^{a}\rangle_{x}A_{a}(\vec{x})=\kappa^{\prime\prime}~.~
\label{constraint gauge}%
\end{equation}
As is usual in applications of the maximum entropy method the specification of
the numerical values of $\kappa^{\prime}$ and $\kappa^{\prime\prime}$ is most
conveniently handled indirectly through the corresponding Lagrange multipliers.

Next we maximize eq.(\ref{entropy a}) subject to (\ref{constraint drift}),
(\ref{constraint gauge}), and normalization. The result is
\begin{equation}
P(x^{\prime}|x)\propto\exp\{-\frac{\alpha}{2}\delta_{ab}\Delta x^{a}\Delta
x^{b}+\alpha^{\prime}\left(  \overline{\partial_{a}\varphi}_{x}-\beta
A_{ax}\right)  \Delta x^{a}\}~ \label{trans prob a}%
\end{equation}
where the Lagrange multipliers $\alpha$, $\alpha^{\prime}$ will be specified
shortly, and the multiplier $\beta$ will eventually be interpreted as the
electric charge, $\beta=q/c$. Alternatively, we can rewrite $P(x^{\prime}|x)$
as
\begin{equation}
P(x^{\prime}|x)=\frac{1}{Z}\exp\left[  -\frac{\alpha}{2}\,\delta_{ab}\left(
\Delta x^{a}-\Delta\overline{x}_{x}^{a}\right)  \left(  \Delta x^{b}%
-\Delta\overline{x}_{x}^{b}\right)  \right]  \label{trans prob b}%
\end{equation}
where
\begin{equation}
\Delta\overline{x}_{x}^{a}=\frac{\alpha^{\prime}}{\alpha}\delta^{ab}\left[
\overline{\partial_{b}\varphi}_{x}-\beta A_{bx}\right]  =\langle\Delta
x^{a}\rangle_{x}~ \label{exp Dx a}%
\end{equation}
is the expected displacement. Using eq.(\ref{spatial evol b}) we can check that%

\begin{align}
\Delta\overline{x}_{x}^{a}  &  =\int dx^{\prime}\,P(x^{\prime}|x)\,\Delta
x^{a}\nonumber\\
&  =%
{\textstyle\sum\nolimits_{k}}
\rho_{k|x}\int dx^{\prime}\,%
{\textstyle\sum\nolimits_{k^{\prime}}}
P(k^{\prime}x^{\prime}|kx)\,\Delta x^{a}=%
{\textstyle\sum\nolimits_{k}}
\rho_{k|x}\langle\Delta x^{a}\rangle_{kx} \label{exp Dxl b}%
\end{align}
is just the spatial displacement averaged over the $k$ variable. (For fixed
$x$ the expectations over $k$ and $x^{\prime}$ commute.)

From \ref{trans prob b}, a generic displacement $\Delta x^{a}$ can be
expressed as the sum of an expected drift, eq.(\ref{exp Dx a}), plus a
fluctuation $\Delta w^{a}$,
\begin{equation}
\Delta x^{a}=x^{\prime a}-x^{a}=\Delta\overline{x}^{a}+\Delta w^{a}\,\,,
\label{Dx}%
\end{equation}
where
\begin{equation}
\left\langle \Delta w^{a}\right\rangle _{x}=0\quad\text{and}\quad\langle\Delta
w^{a}\Delta w^{b}\rangle_{x}=\frac{1}{\alpha}\delta^{ab}~. \label{fluct a}%
\end{equation}

\section{Entropic time: instants and duration}

\label{entropic time}In ED time is introduced as a book-keeping device
designed to keep track of the accumulation of short steps (\cite{Caticha
2019})(\cite{Caticha 2024}). It involves identifying suitable notions of
ordered \textquotedblleft instants\textquotedblright\ and of the separation or
duration between \textquotedblleft successive\textquotedblright\ instants.
Just as the prototype of a classical clock is a free particle that
\textquotedblleft measures\textquotedblright\ equal intervals by registering
equal displacements, the prototype of a quantum clock is the quantum system
itself. This implies that in order to recover a notion of time as a continuous
succession of instants it is necessary to appeal to the continuity of spatial trajectories.

Referring to the discussion in the previous section, specifying the duration
or interval $\Delta t=t^{\prime}-t$ between successive instants amounts to
specifying the relation between $\Delta t$ and the multipliers $\alpha$ and
$\alpha^{\prime}$ \cite{Caticha 2019}. The basic criterion is convenience:
\textquotedblleft duration is defined so that motion looks
simple\textquotedblright. To reflect the translational symmetry of a
non-relativistic Newtonian space and time we choose $\alpha^{\prime}$ and
$\alpha$ to be independent of $x$ and $t$ so that time flows \textquotedblleft
equably everywhere and everywhen.\textquotedblright\ We \emph{define} $\Delta
t$ so that $\alpha^{\prime}/\alpha\propto\Delta t$. Then, as we see from
(\ref{exp Dx a}), the particle has a well defined expected velocity. For later
convenience the proportionality constant is written as $1/m$,
\begin{equation}
\frac{\alpha^{\prime}}{\alpha}=\frac{1}{m}\Delta t~. \label{alphas}%
\end{equation}
At this point the constant $m$ has no special significance but, once we derive
the Pauli equation, it will be recognized as the particle's mass. It remains
to specify $\alpha$. We choose $\alpha$ so that for sufficiently short steps
the expected fluctuations, $\langle\Delta w^{a}\Delta w^{b}\rangle_{x}$,
increase by equal amounts in equal intervals $\Delta t$. Referring to
eq.(\ref{fluct a}) this is achieved by setting
\begin{equation}
\frac{1}{\alpha}=\frac{\eta}{m}\Delta t\quad\text{so that}\quad\alpha^{\prime
}=\frac{1}{\eta}~. \label{alpha prime a}%
\end{equation}
where a new constant $\eta$ is introduced. We emphasize that these choices of
$\alpha$ and $\alpha^{\prime}$ are not arbitrary as they lead to a natural
physical interpretation: duration is defined so it reflects the symmetries of
Newtonian space and time, and so that over short steps particles have well
defined expected velocities while equal intervals of entropic time correspond
to equal increases of the variance $\langle\Delta w^{2}\rangle$. Indeed,
substituting (\ref{alpha prime a}) into eqs.(\ref{exp Dx a}) and
(\ref{fluct a}) we find that a generic displacement is%

\begin{equation}
\Delta x^{a}=\Delta\overline{x}_{x}^{a}+\Delta w^{a}=b_{x}^{a}\Delta t+\Delta
w^{a}~, \label{delta x b}%
\end{equation}
where
\begin{equation}
b_{x}^{a}=\frac{\langle\Delta x^{a}\rangle_{x}}{\Delta t}=\frac{1}{m}%
\delta^{ab}\left[  \overline{\partial_{a}\varphi}_{x}-\beta A_{bx}\right]  ~,
\label{drift velocity a}%
\end{equation}
is the drift velocity, and the spatial fluctuations $\Delta w^{a}$ obey
\begin{equation}
\left\langle \Delta w^{a}\right\rangle _{x}=0\quad\text{and}\quad\langle\Delta
w^{a}\Delta w^{b}\rangle_{x}=\frac{\eta}{m}\delta^{ab}\Delta t~.
\label{ED fluct b}%
\end{equation}

We are now ready to investigate the consequences of the spatial transition
probability,
\begin{equation}
P(x^{\prime}|x)=\frac{1}{Z}\exp\left[  -\frac{m}{2\eta\Delta t}\,\delta
_{ab}\left(  \Delta x^{a}-\Delta\overline{x}_{x}^{a}\right)  \left(  \Delta
x^{b}-\Delta\overline{x}_{x}^{b}\right)  \right]  ~, \label{trans prob c}%
\end{equation}
by rewriting the equation of evolution for the spatial distribution $\rho(x)$,
eq.(\ref{spatial evol b}), as a differential equation.

\section{The probability evolution equation}

\label{Probability evolution}Multiply eq.(\ref{spatial evol b}) by a smooth
test function $f_{x^{\prime}}$ and integrate over $x^{\prime}$,
\begin{equation}
\int dx^{\prime}\,\rho_{x^{\prime}}^{\prime}f_{x^{\prime}}=\int dx^{\prime
}\int dx\,P(x^{\prime}|x)\rho_{x}f_{x^{\prime}}=\int dx\left[  \int
dx^{\prime}P(x^{\prime}|x)\,f_{x^{\prime}}\right]  \rho_{x}~.{} \label{CKc}%
\end{equation}
The test function $f_{x^{\prime}}$ is assumed sufficiently smooth precisely so
that it can be expanded about $x$. Terms $(\Delta x)^{2}$ contribute to
$O(\Delta t)$. Then, dropping all terms of order higher than $\Delta t$, the
integral in the brackets is
\begin{align}
\left[  \cdots\right]   &  =\int dx^{\prime}\,P(x^{\prime}|x)\left(
f_{x}+\frac{\partial f_{x}}{\partial x^{a}}\Delta x^{a}+\frac{1}{2}%
\frac{\partial^{2}f_{x}}{\partial x^{a}\partial x^{b}}\Delta x^{a}\Delta
x^{b}+\ldots\right) \nonumber\\
&  =f_{x}+b_{x}^{a}\Delta t\frac{\partial f}{\partial x^{a}}+\frac{1}{2}\Delta
t\,\frac{\eta}{m}\delta^{ab}\frac{\partial^{2}f}{\partial x^{a}\partial x^{b}%
}+\ldots\label{CKd}%
\end{align}
where we used eqs.(\ref{drift velocity a}) and (\ref{ED fluct b}). Dropping
the primes on the left hand side of (\ref{CKc}), substituting (\ref{CKd}) into
the right, and dividing by $\Delta t$, gives
\begin{equation}
\int dx\,\frac{1}{\Delta t}\left[  \rho_{x}^{\prime}-\rho_{x}\right]
f_{x}=\int dx\left[  b_{x}^{a}\frac{\partial f_{x}}{\partial x^{a}}+\frac
{1}{2}\,\frac{\eta}{m}\delta^{ab}\frac{\partial^{2}f_{x}}{\partial
x^{a}\partial x^{b}}\right]  \rho_{x}~.{}%
\end{equation}
Next integrate by parts on the right and let $\Delta t\rightarrow0$. Since the
test function $f(x)$ is arbitrary, we conclude that
\begin{equation}
\partial_{t}\rho=-\partial_{a}(b^{a}\rho)+\frac{1}{2}\frac{\eta}{m}\delta
^{ab}\partial_{a}\partial_{b}\rho~.~ \label{FP a}%
\end{equation}
This can be written in the alternative form,
\begin{equation}
\partial_{t}\rho_{t}=-\partial_{a}\left(  v_{x}^{a}\rho_{x}\right)  ~,
\label{prob evol c}%
\end{equation}
which is a continuity equation, where%
\begin{equation}
v_{x}^{a}=b_{x}^{a}-\frac{\eta}{2m}\delta^{ab}\partial_{b}\log\rho_{x}
\label{current vel a}%
\end{equation}
is the \textquotedblleft current\textquotedblright\ velocity of the
probability flow.

\section{The e-Hamiltonian and the Pauli equation}

\label{Hamiltonian}The whole purpose of this exercise has been to formulate
the physical fact that particle paths are continuous in a way that allows us
to identify a suitable notion of time. It is natural to require that
acceptable Hamiltonians be such that they generate evolution along a time
defined in terms of the very same \textquotedblleft clock\textquotedblright%
\ (the system itself) that provides the measure of time. Therefore, our
immediate goal is to identify those Hamiltonians $\tilde{H}$ that
reproduce\ the continuity equation (\ref{prob evol c}). On the other hand,
using eqs.(\ref{rho csi}) and (\ref{sympl form b}) we can write $\partial
\rho_{x}/\partial t$ directly in Hamiltonian form. Since $(\rho_{kx},\xi
_{kx})$ are canonically conjugate, then
\begin{equation}
\rho_{x}=\rho_{+x}+\rho_{-x}\quad\text{and}\quad\xi_{x}=\frac{1}{2}\left(
\xi_{+x}+\xi_{-x}\right)  ~ \label{rho csi b}%
\end{equation}
are canonically conjugate too\ (we can check that $\{\rho_{x},\xi_{x^{\prime}%
}\}=\delta_{xx^{\prime}}$) which leads to
\begin{equation}
\frac{\partial\rho_{x}}{\partial t}=\frac{\partial\rho_{+x}}{\partial t}%
+\frac{\partial\rho_{-x}}{\partial t}=\frac{\delta\tilde{H}}{\delta\xi_{+x}%
}+\frac{\delta\tilde{H}}{\delta\xi_{-x}}=\frac{\delta\tilde{H}}{\delta\xi_{x}%
}~. \label{prob evol d}%
\end{equation}
As we see from eqs.(\ref{prob evol c}) and (\ref{prob evol d}), at this point
in our argument we have two independent expressions for $\partial\rho
_{x}/\partial t$ and, therefore, there must exist some intimate connection
between the drift potentials $\varphi_{kx}$ and the phases $\xi_{kx}$. To find
what this relation might be, examine eqs.(\ref{drift velocity a}) and
(\ref{current vel a}) and rewrite the current velocity as%

\begin{equation}
v_{x}^{a}=\frac{1}{m}\delta^{ab}\partial_{b}%
{\textstyle\sum\limits_{k}}
\rho_{k|x}\left(  \varphi_{kx}-\frac{\eta}{2}\log\rho_{x}\right)  -\frac
{\beta}{m}A_{x}^{a}~.
\end{equation}
We propose that the desired relation is%

\begin{equation}
\xi_{kx}=\varphi_{kx}-\eta\log\rho_{x}^{1/2}~, \label{can transf a}%
\end{equation}
so that
\begin{equation}
v_{x}^{a}=\frac{\delta^{ab}}{m}\left(
{\textstyle\sum\limits_{k}}
\rho_{k|x}\partial_{b}\xi_{kx}-\beta A_{bx}\right)  ~. \label{current vel b}%
\end{equation}
We note that eq.(\ref{can transf a}) is a simple canonical transformation;
since $\varphi_{kx}$ and $\xi_{kx}$ differ by a function of the generalized
coordinate $\rho_{kx}$ they are equally legitimate choices of canonical
momenta --- the choice of $\xi_{kx}$ over $\varphi_{kx}$ is purely a matter of convenience.

The other piece of information guiding our choice of e-Hamiltonian is the fact
that for HK flows the e-Hamiltonian is bilinear in the wave functions,
eq.(\ref{e-hamiltonian a}). The next step, therefore, is to write
eqs.(\ref{prob evol c}) and (\ref{current vel b}) in terms of the wave
functions, $\psi_{kx}=\rho_{kx}^{1/2}\exp\frac{i}{\hbar}\xi_{kx}$. After a
little algebra we find
\begin{equation}
\frac{\partial\rho_{x}}{\partial t}=\frac{\hbar i}{2m}%
{\textstyle\sum\limits_{k}}
\partial_{a}\left[  \psi_{k}^{\ast}\left(  D^{a}\psi_{k}\right)  -\psi
_{k}\left(  D^{a}\psi_{k}\right)  ^{\ast}\right]  ~,
\end{equation}
where $D_{a}=\partial_{a}-\frac{iq}{\hbar c}A_{a}$ is the covariant
derivative. Further rearranging yields
\begin{equation}
\frac{\partial\rho_{x}}{\partial t}=\frac{\hbar}{2mi}\left[  \psi_{k}\left(
D_{a}D_{a}\psi_{k}\right)  ^{\ast}-\psi_{k}^{\ast}D_{a}D_{a}\psi_{k}\right]
=\frac{\delta\tilde{H}}{\delta\xi_{x}}~, \label{prob evol e}%
\end{equation}
which is a \emph{linear} functional differential equation for $\tilde{H}$ that
is easily integrated,
\begin{equation}
\tilde{H}[\psi,\psi^{\ast}]=\frac{-\hbar^{2}}{2m}\int dx\,\,%
{\textstyle\sum\nolimits_{k}}
\psi_{kx}^{\ast}D_{a}D_{a}\psi_{kx}+\tilde{V}~, \label{e-hamiltonian b}%
\end{equation}
where the integration constant $\tilde{V}$ is independent of $\xi_{x}$. This
is easily checked: let $\xi_{x}\rightarrow\xi_{x}+\delta\xi_{x}$ and use
\begin{equation}
\delta\psi_{kx}=\frac{i}{\hbar}\psi_{kx}\,\delta\xi_{x}\quad\text{and}%
\quad\delta\psi_{kx}^{\ast}=-\frac{i}{\hbar}\psi_{kx}^{\ast}\,\delta\xi_{x}~,
\end{equation}
to get
\begin{equation}
\delta\tilde{H}=\frac{\hbar}{2mi}\int dx\,\,%
{\textstyle\sum\nolimits_{k}}
\left[  \psi_{kx}\left(  D_{a}D_{a}\psi_{kx}\right)  ^{\ast}-\psi_{kx}^{\ast
}D_{a}D_{a}\psi_{kx}\right]  \delta\xi_{x}\qquad\text{qed.}%
\end{equation}
Thus we see that the kinetic energy in the e-Hamiltonian
(\ref{e-hamiltonian b}) is traced to the entropic updating that led to the
spatial continuity equation, while the potential energy is introduced as an
integration constant. To determine $\tilde{V}$ we note that in order for
$\tilde{H}$ to generate an HK flow we must require that $\tilde{V}$ be
bilinear in $\psi$,%

\begin{equation}
\tilde{V}[\psi,\psi^{\ast}]=\int dxdx^{\prime}%
{\textstyle\sum\limits_{kk^{\prime}}}
\psi_{kx}^{\ast}\hat{V}_{kx,k^{\prime}x^{\prime}}\psi_{k^{\prime}x^{\prime}}
\label{V a}%
\end{equation}
for some Hermitian kernel $\hat{V}_{kx,k^{\prime}x^{\prime}}$ and,
furthermore, to reproduce the ED evolution in eq.(\ref{prob evol e}),
$\tilde{V}$ must be independent of $\xi_{x}=(\xi_{+x}+\xi_{-x})/2$,
\begin{equation}
\frac{\delta\tilde{V}}{\delta\xi_{x}}=\frac{\delta}{\delta\xi_{x}}\int
dxdx^{\prime}%
{\textstyle\sum\limits_{kk^{\prime}}}
\rho_{kx}^{1/2}\rho_{k^{\prime}x^{\prime}}^{1/2}\hat{V}_{kx,k^{\prime
}x^{\prime}}\exp\frac{i}{\hbar}(\xi_{k^{\prime}x^{\prime}}-\xi_{kx})=0~.
\label{V b}%
\end{equation}
To satisfy (\ref{V b}) for \emph{arbitrary} choices of $\rho_{kx}$ and
$\rho_{k^{\prime}x^{\prime}}$ we require that the kernel $\hat{V}%
_{kx,k^{\prime}x^{\prime}}$ be \emph{local} in $x$,
\begin{equation}
\hat{V}_{kx,k^{\prime}x^{\prime}}=\delta_{xx^{\prime}}\hat{V}_{kk^{\prime}%
}(x)~,
\end{equation}
The local Hermitian kernel $\hat{V}_{kk^{\prime}}(x)$ is a Hermitian
$2\times2$ matrix which can be expanded in terms of Pauli matrices,%
\begin{equation}
\hat{V}_{kk^{\prime}}(x)=V_{0}(x)\delta_{kk^{\prime}}+V_{a}(x)\sigma
_{kk^{\prime}}^{a}~,\quad(a=1,2,3)~,
\end{equation}
where $V_{0}(x)$ and $V_{a}(x)$ are four scalar functions,
\begin{equation}
\tilde{V}[\psi,\psi^{\ast}]=\int dx\left(  V_{0x}%
{\textstyle\sum\limits_{k}}
\psi_{kx}^{\ast}\psi_{kx}+V_{ax}%
{\textstyle\sum\limits_{kk^{\prime}}}
\psi_{kx}^{\ast}\sigma_{kk^{\prime}}^{a}\psi_{k^{\prime}x}\right)
\end{equation}
(and we can easily check that, indeed, $\delta\tilde{V}/\delta\xi_{x}=0$). The
final expression for the e-Hamiltonian
\begin{equation}
\tilde{H}=\int dx\,%
{\textstyle\sum\limits_{kk^{\prime}}}
\,\psi_{kx}^{\ast}\left[  \left(  \frac{-\hbar^{2}}{2m}D_{a}D_{a}%
+V_{0}\right)  \delta_{kk^{\prime}}+V_{a}\sigma_{kk^{\prime}}^{a}\right]
\psi_{k^{\prime}x}~, \label{e-hamiltonian c}%
\end{equation}
takes the form of an expected value. The corresponding Schr\"{o}dinger
equation --- the Pauli equation --- is%
\begin{equation}
i\hbar\frac{\partial\psi_{kx}}{\partial t}=\frac{\delta\tilde{H}}{\delta
\psi_{kx}^{\ast}}=\frac{-\hbar^{2}}{2m}\delta^{ab}D_{a}D_{b}\psi_{kx}%
+V_{0x}\psi_{kx}+V_{ax}\sigma_{kk^{\prime}}^{a}\psi_{k^{\prime}x}~.
\label{Pauli eq a}%
\end{equation}
Incidentally, at this stage in the development we see that the constants
$\hbar$, $m$, and $q/c$ can be safely given their usual meanings, and also
that the entropic time $t$ is the actual time measured by clocks --- after
all, it is by using equations such as the Schr\"{o}dinger or the Pauli
equations that we calibrate our measuring devices.

For a \emph{point} particle, such as an electron, the Pauli equation takes its
simplest form,%
\begin{equation}
i\hbar\partial_{t}\Psi_{x}=\frac{1}{2m}\left(  \frac{\hbar}{i}\vec{\partial
}-\frac{q}{c}\vec{A}\right)  ^{2}\Psi_{x}+qA_{0}\Psi_{x}-\frac{\hbar q}%
{2mc}B_{a}\sigma^{a}\Psi_{x}~, \label{Pauli eq c}%
\end{equation}
or
\begin{equation}
i\hbar\partial_{t}\Psi_{x}=\frac{1}{2m}\left[  \hat{\sigma}^{a}\left(
\frac{\hbar}{i}\partial_{a}-\frac{q}{c}A_{a}\right)  \right]  ^{2}\Psi
_{x}+qA_{0}\Psi_{x}~, \label{Pauli eq b}%
\end{equation}
where $\Psi_{x}$ is given in eq.(\ref{spinor a}) and
\begin{equation}
V_{0x}=qA_{0}(x)\quad\text{and}\quad V_{ax}=-\frac{\hbar q}{2mc}B_{a}(x)~
\end{equation}
represent the interactions of the charge with the scalar electric potential
$A_{0}(x)$ and of the magnetic dipole with the magnetic field $B_{a}$.
Equation (\ref{Pauli eq a}) is more general in that it could describe extended
particles such as protons and neutrons with anomalous magnetic moments, or
even particles with electric dipole moments. (The latter would violate
time-reversal invariance and have not been observed so far.)

One may note that Hilbert spaces have not been mentioned; strictly, they are
not needed. However, the linearity of eq.(\ref{Pauli eq a}) suggests that a
Hilbert space is a convenient calculational tool \cite{Caticha 2019}%
\cite{Caticha 2024}. Here, for completeness, we briefly mention how to
introduce vectors $|\Psi\rangle$ in a linear Hilbert space. The map to the
Dirac notation, $\psi_{kx}\leftrightarrow|\Psi\rangle$, is defined by
\begin{equation}
|\Psi\rangle=\int dx%
{\textstyle\sum\limits_{k}}
\,|kx\rangle\psi_{kx}\quad\text{where}\quad\psi_{kx}=\langle kx|\Psi\rangle~,
\end{equation}
where, in this \textquotedblleft$kx$\textquotedblright\ representation,\ the
vectors $\{|kx\rangle\}$ form a basis that is orthogonal and complete,
\begin{equation}
\langle kx|k^{\prime}x^{\prime}\rangle=\delta_{kk^{\prime}}\delta_{xx^{\prime
}}\quad\text{and}\quad\int dx%
{\textstyle\sum\limits_{k}}
\,|kx\rangle\langle kx|~=\hat{1}~.
\end{equation}
The Hilbert scalar product $\langle\Psi_{1}|\Psi_{2}\rangle$ is then defined
by exploiting the structures already available to us, eqs.(\ref{Psi mu}),
(\ref{sympl form c}), and (\ref{spinor metric e}) for $\Psi^{\mu x}$, the
symplectic form $\Omega$ and the metric $G$,
\begin{equation}
\langle\Psi_{1}|\Psi_{2}\rangle\overset{\text{def}}{=}\frac{1}{2\hbar}\int
dxdx^{\prime}%
{\textstyle\sum\limits_{\mu\mu^{\prime}}}
\left(  G_{\mu x,\mu^{\prime}x^{\prime}}+i\Omega_{\mu x,\mu^{\prime}x^{\prime
}}\right)  \Psi_{1}^{\mu x}\Psi_{2}^{\mu^{\prime}x^{\prime}}~.
\end{equation}
A straightforward calculation yields the familiar expression
\begin{equation}
\langle\Psi_{1}|\Psi_{2}\rangle=\int dx\,%
{\textstyle\sum\limits_{k}}
\psi_{1kx}^{\ast}\psi_{2kx}~.
\end{equation}
The e-Hamiltonian is given by the expected value,
\begin{equation}
\tilde{H}=\langle\Psi|\hat{H}|\Psi\rangle\quad\text{with}\quad\hat
{H}_{kx,k^{\prime}x^{\prime}}=\langle kx|\hat{H}|k^{\prime}x^{\prime}\rangle~,
\end{equation}
where the matrix element $\hat{H}_{kx,k^{\prime}x^{\prime}}$ can be read off
eq.(\ref{e-hamiltonian c}).

\section{Orbital and spin angular momenta \ }

\label{angular momentum}We finally come to the \textquotedblleft
reconstruction\textquotedblright\ of spin. The central input is that angular
momentum is the generator of rotations. Under a rotation $\vec{\zeta}%
=\zeta\vec{n}$ by an infinitesimal angle $\zeta$ about the axis $\vec{n}$ we
have ,
\begin{equation}
\vec{x}\rightarrow\vec{x}_{\zeta}=\vec{x}+\zeta\vec{n}\times\vec{x}~.
\end{equation}
The action of rotations on the \emph{spatial} probability distribution
$\rho(x)$ is given by%
\begin{equation}
\rho(x)\rightarrow\rho_{\zeta}(x)\quad\text{with}\quad\rho_{\zeta}(x_{\zeta
})=\rho(x)\quad\text{or}\quad\rho_{\zeta}(x)=\rho(x_{-\zeta})~.
\end{equation}
so that%
\begin{equation}
\delta_{\zeta}\rho(x)=\rho_{\zeta}(x)-\rho(x)=-\zeta\vec{n}\cdot\vec{x}%
\times\vec{\partial}\rho_{x}=-\varepsilon^{abc}\zeta n_{a}x_{c}\partial
_{c}\rho_{x}~.
\end{equation}
Therefore
\begin{equation}
\frac{\partial\rho_{x}}{\partial\zeta}=-\varepsilon^{abc}n_{a}x_{b}%
\partial_{c}\rho_{x}~. \label{rot flow a}%
\end{equation}
On the other hand, since rotations about $n^{a}$ are generated by the
Hamiltonian function $\tilde{J}^{a}n_{a}$, and as we saw in
eq.(\ref{rho csi b}) $\{\rho_{x},\xi_{x}\}$ are canonically conjugate, we can
write the Hamilton equation
\begin{equation}
\frac{\partial\rho_{x}}{\partial\zeta}=\frac{\partial\rho_{+x}}{\partial
t}+\frac{\partial\rho_{-x}}{\partial t}=\frac{\delta\tilde{J}^{a}}{\delta
\xi_{+x}}n_{a}+\frac{\delta\tilde{J}^{a}}{\delta\xi_{-x}}n_{a}=\frac
{\delta\tilde{J}^{a}}{\delta\xi_{x}}n_{a}\ . \label{rot flow c}%
\end{equation}
Combining with (\ref{rot flow a}) leads to
\begin{equation}
-\varepsilon^{abc}n_{a}x_{c}\partial_{c}\rho_{x}=\frac{\delta\tilde{J}^{a}%
}{\delta\xi_{x}}n_{a}~, \label{rot flow b}%
\end{equation}
which is a linear differential equation for $\tilde{J}^{a}n_{a}$. The
integration is easy and since (\ref{rot flow b}) holds for any choice of
$n^{a}$, we find%
\begin{equation}
\tilde{J}^{a}=\int dx\,\,%
{\textstyle\sum\nolimits_{k}}
\psi_{kx}^{\ast}\left(  \varepsilon^{abc}x_{b}\frac{\hbar}{i}\partial
_{c}\right)  \psi_{kx}+\tilde{S}^{a}~. \label{ang mom b}%
\end{equation}
The first integral is recognized as orbital angular momentum, and $\tilde
{S}^{a}$ is an integration constant independent of $\xi_{x}$, which we
identify as the \emph{spin functional}. (Eq.(\ref{ang mom b}) is easy to
check: just let $\xi_{x}\rightarrow\xi_{x}+\delta\xi_{x}$ with $\delta
\psi_{kx}=i\psi_{kx}\,\delta\xi_{x}/\hbar$, then integrate by parts using
$\varepsilon^{abc}\partial_{c}x^{b}=0$.)

To determine the spin functional $\tilde{S}^{a}$ we impose two natural
conditions: The first, mentioned in section \ref{HK flows}, is that $\tilde
{J}^{a}n_{a}$ is a Hamiltonian function --- it generates an HK flow. The
second condition is motivated by the choice of ontology: the two-valuedness of
$k$ requires that the action of $\tilde{S}^{a}$ on the wave functions given by
eq.(\ref{spinor a}) results in a \emph{2-dimensional} representation of the
rotation group.

Concerning the first condition: in order for $\tilde{J}^{a}$ in
(\ref{ang mom b}) to generate an HK flow we require that $\tilde{S}^{a}$ be
also bilinear in $\psi$,%

\begin{equation}
\tilde{S}^{a}[\psi,\psi^{\ast}]=\int dxdx^{\prime}%
{\textstyle\sum\limits_{kk^{\prime}}}
\psi_{kx}^{\ast}\hat{S}_{kx,k^{\prime}x^{\prime}}^{a}\psi_{k^{\prime}%
x^{\prime}}~,
\end{equation}
for some Hermitian kernel $\hat{S}_{kx,k^{\prime}x^{\prime}}^{a}$.
Furthermore, to reproduce eq.(\ref{rot flow b}), $\hat{S}_{kx,k^{\prime
}x^{\prime}}^{a}$ must be independent of $\xi_{x}=(\xi_{+x}+\xi_{-x})/2$,
\begin{equation}
0=\frac{\delta\tilde{S}^{a}}{\delta\xi_{x}}=\frac{\delta}{\delta\xi_{x}}\int
dxdx^{\prime}%
{\textstyle\sum\limits_{kk^{\prime}}}
\rho_{kx}^{1/2}\rho_{k^{\prime}x^{\prime}}^{1/2}\hat{S}_{kx,k^{\prime
}x^{\prime}}^{a}\exp\frac{i}{\hbar}(\xi_{k^{\prime}x^{\prime}}-\xi_{kx})~,
\end{equation}
for \emph{arbitrary} choices of $\rho_{kx}$ and $\rho_{k^{\prime}x^{\prime}}$.
It follows that $\hat{S}_{kx,k^{\prime}x^{\prime}}^{a}$ must be local in $x$,
\begin{equation}
\hat{S}_{kx,k^{\prime}x^{\prime}}^{a}=\delta_{xx^{\prime}}\hat{S}_{kk^{\prime
}}^{a}(x)\quad\text{or}\quad\tilde{S}^{a}=\int dx%
{\textstyle\sum\limits_{kk^{\prime}}}
\psi_{kx}^{\ast}\hat{S}_{kk^{\prime}}^{a}(x)\psi_{k^{\prime}x}~.
\label{spin funct a}%
\end{equation}

To implement the second condition we recall that a 2-dimensional
representation of rotations in terms of Cayley-Klein parameters was already
known before the discovery of QM (see, \emph{e.g.,}\cite{Goldstein 2000}). In
a $2\times2$ matrix representation a generic vector $\vec{x}$ is represented
by
\begin{equation}
\vec{x}=x^{a}\sigma_{a}=%
\begin{bmatrix}
x^{3} & x^{1}-ix^{2}\\
x^{1}+ix^{2} & -x^{3}%
\end{bmatrix}
~,
\end{equation}
where the basis vectors $\vec{e}_{a}$ are represented by Pauli matrices
$\sigma_{a}$, and a rotation by $\zeta$ about the axis $\vec{n}$ is
represented by
\begin{equation}
\vec{x}^{\prime}=\hat{U}_{\zeta}\vec{x}\hat{U}_{\zeta}^{-1}\quad
\text{where}\quad\hat{U}_{\zeta}=\exp(-in^{a}\sigma_{a}\zeta/2)~.
\end{equation}
So far this has nothing to do with QM but the $2\times2$ matrices $\hat
{U}_{\zeta}$ form a group and the corresponding rotation of the wave function
$\Psi$ is given by
\begin{equation}
\Psi(x)\rightarrow\Psi_{\zeta}(x_{\zeta})=\hat{U}_{\zeta}\Psi(x)\quad
\text{or}\quad\Psi_{\zeta}(x)=\hat{U}_{\zeta}\Psi(x_{-\zeta})~,
\end{equation}
which suggests we rewrite
\begin{equation}
\hat{U}_{\zeta}=\exp\left(  -i\hat{S}^{a}\zeta/\hbar\right)  \quad
\text{with}\quad\hat{S}_{kk^{\prime}}^{a}=\frac{\hbar}{2}\sigma_{kk^{\prime}%
}^{a}~.~
\end{equation}
From (\ref{spin funct a}) the spin functional we seek is%

\begin{equation}
\tilde{S}^{a}[\psi,\psi^{\ast}]=\int dx%
{\textstyle\sum\limits_{kk^{\prime}}}
\psi_{kx}^{\ast}\frac{\hbar}{2}\sigma_{kk^{\prime}}^{a}\psi_{k^{\prime}x}~.
\label{spin funct b}%
\end{equation}
The translation to Hilbert space in Dirac notation is, once again,
straightforward. For example, the spin functional is the expected value,
\begin{equation}
\tilde{S}^{a}=\langle\Psi|\hat{S}^{a}|\Psi\rangle\quad\text{with}\quad\hat
{S}_{kx,k^{\prime}x^{\prime}}^{a}=\langle kx|\hat{S}^{a}|k^{\prime}x^{\prime
}\rangle~,
\end{equation}
where the matrix element $\hat{S}_{kx,k^{\prime}x^{\prime}}^{a}$ can be read
off eq.(\ref{spin funct b}). This concludes the reconstruction of spin.

\section{Discussion and conclusions}

\label{conclusions} We have reconstructed the mathematical formalism for the
QM of a spin-1/2 particle and recovered the linearity of the Pauli equation,
the emergence of complex numbers, the peculiar properties of spin 1/2, and
more. This guarantees that the ED predictions are in complete agreement with
experiments, which is a feature that ED shares with so many other
interpretations of QM. Where ED claims an additional dose of success is in
achieving ontological and epistemic clarity. On the ontological side, it
provides a crisp answer to the question \textquotedblleft what is
real?\textquotedblright: position and \textquotedblleft$k$%
-ness\textquotedblright\ are real. ED affords clear epistemic status to
probabilities, wave functions, energies, and angular momenta, including spin,
and does this while enlisting the proven methods of modern quantitative
epistemology, namely entropic methods and information geometry, without the
need to invoke exotic or quantum probabilities. In this latter sense ED is a
rather conservative model. However, ED turns out to be radically non-classical
in that it is a dynamics of probabilities and not of particles. ED denies the
ontic status of dynamics and it is the latter aspect of ED that violates our
classical intuitions. To be clear, if at one instant probabilities are large
in one place, and at a later instant they are large somewhere else, we are
correct to believe that the particles have moved --- but nothing in ED
describes the causal mechanism that pushed the particles around. As far as we
can tell, there is none; ED is a \emph{mechanics without a mechanism}.

Now that the dynamics has been fully developed we can revisit some questions
raised back in Section \ref{intro}. What is this mysterious, not classically
describable $k$? What is spin? How are $k$ and spin related? What, if
anything, is spinning?

Not much can be said about $k$ except that it takes two values and it belongs
in the ontic sector. Perhaps this is just the way the world is and it is all
one can ever hope to say. Concerning spin, much has been said in the previous
section: it belongs in the epistemic sector, it generates HK flows and
rotations, and much of its spinor structure follows from the two-valuedness of
$k$.

The relation between $k$ and spin is not without interest. The important
difference is that the former is ontic and the latter epistemic --- as
different as different can be. However, there tends to exist a 1-1
correspondence between the ontic microstates ($k=+1$ and $k=-1$) and the
epistemic states that represent certainty about them (the spinors $\binom
{1}{0}$ and $\binom{0}{1}$) and this might be a source of confusion. But some
mystery is bound to remain. Consider a system in the epistemic state described
by $\binom{1}{0}$; we are certain that $k=+1$. Suppose the system is then
rotated by $\theta=\pi/2$ about the $\vec{e}_{y}$ axis,
\begin{equation}
\exp\left(  -i\sigma_{y}\frac{\pi/2}{2}\right)  \binom{1}{0}=\frac{1}{\sqrt
{2}}\binom{1}{1}~.
\end{equation}
The outcome of this physical operation is that now we ought to believe that
there is a 50\% probability that a transition from $k=1$ to $k=-1$ has
occurred. ED is silent about what could have caused this transition to happen.
The weirdness of a \emph{mechanics without a mechanism} can manifest itself
not just in dynamics but also in the context of rotations or other operations.

Finally, there is the question of what is spinning? The assertions that
probability, energy, momentum, and angular momentum including spin are not
ontic but epistemic quantities force an extreme revision of our intuitions
about physics. Probabilities may change but they neither move nor flow; they
are not substances. Similarly, accepting that spin is an epistemic concept
forces us to revisit the intuition that something substantial is actually
spinning. ED is just a model that describes a fictitious world. The real world
out there may contain all sorts of things, but the fictitious world described
in this paper contains a single point particle characterized by its position
and its discrete $k$ value, that's all. Within this ED model the answer is
clear: there is nothing there that could spin and, therefore, \emph{nothing is
spinning}.

The question, of course, is whether the ED models provide successful guidance
in the real world, and not just in their own fictitious worlds. So far,
everything indicates they do.

\subparagraph*{Acknowledgments}

I would like to thank N. Carrara, S. Nawaz, and K. Vannier for valuable
discussions on the entropic dynamics of spin and their contributions at
various stages of this program. I would also like to thank F. Holik for a
useful exchange about the Pauli Exclusion Principle at the X Conference on
Quantum Foundations in Buenos Aires, Argentina, 2021.

\end{document}